\documentclass[useAMS,usenatbib]{mn2e} 
\usepackage{xcolor}
\usepackage{aas_macros}
\usepackage{graphics}
\usepackage[pdftex]{graphicx}
\usepackage{epstopdf}
\usepackage{epsfig}  
\usepackage{natbib} 
\usepackage{dingbat}
\usepackage{float}
\usepackage{amsmath}
\usepackage{times}
\usepackage[varg]{txfonts}
\usepackage{verbatim} 
\usepackage{multirow,bigdelim} 
\usepackage{color}
\usepackage{vmargin}
\usepackage{array,booktabs,arydshln,xcolor}

\setmarginsrb{1.2cm}{1cm}{1.2cm}{1cm}{1cm}{1cm}{1cm}{1cm}

\newcommand{\hmsun}{{\, h^{-1}\rm~M}_\odot}

  \makeatletter
    \renewcommand{\paragraph}{\@startsection{paragraph}{4}{\z@}%
      {-3.25ex\@plus -1ex \@minus -.2ex}%
      {1.5ex \@plus .2ex}%
      {\normalfont\small\centering}}
     
    \renewcommand{\subparagraph}{\@startsection{subparagraph}{5}{\z@}%
      {-3.25ex\@plus -1ex \@minus -.2ex}%
      {1.5ex \@plus .2ex}%
      {\normalfont\small\centering}}
    \makeatother

\newcommand{\rockstar}{{\sc Rockstar}}
\newcommand{\music}{{\sc music}}

\newcommand{\hMpc}{{ \textit{h}$^{-1}$~Mpc}}

\title[History \& Neighbors]{The Past History of Galaxy Clusters told by their present neighbors}
\author[Sorce et al.]
{{Jenny G. Sorce$^{1,2,3}$\thanks{E-mail: \text{jenny.sorce@univ-lyon1.fr / jenny.sorce@ens-lyon.fr / jsorce@aip.de}}, 
Stefan Gottl\"ober$^{3}$, Gustavo Yepes$^{4,5}$
}\\
$^1$Univ Lyon, ENS de Lyon, Univ Lyon1, CNRS, Centre de Recherche Astrophysique de Lyon UMR5574, F-69007, Lyon, France\\
$^2$Univ Lyon, Univ Lyon1, ENS de Lyon, CNRS, Centre de Recherche Astrophysique de Lyon UMR5574, F-69230, Saint-Genis-Laval, France\\
$^3$Leibniz-Institut f\"{u}r Astrophysik (AIP), An der Sternwarte 16, D-14482 Potsdam, Germany\\
$^{4}$Departamento de F\'{\i}sica Te\'orica, Universidad Aut\'onoma de Madrid, Cantoblanco E-28049, Madrid, Spain \\ 
$^{5}$Centro de Investigaci\'on Avanzada en F\'{\i}sica Fundamental,  Facultad de Ciencias, Universidad Aut\'onoma de Madrid, E-28049 Madrid, Spain \\
}

\begin{document}

\date{}

\pagerange{\pageref{firstpage}--\pageref{lastpage}} \pubyear{2020}

\maketitle

\label{firstpage}

\begin{abstract}
\indent 

 Galaxy clusters can play a key role in modern cosmology provided their evolution is properly understood. However, observed clusters give us only a single timeframe of their dynamical state. Therefore, finding present observable data of clusters that are well correlated to their assembly history constitutes an inestimable tool for cosmology. Former studies correlating  environmental descriptors of clusters to their formation history are dominated by halo mass - environment relations. This paper presents a mass-free correlation between the present neighbor distribution of cluster-size halos and the latter mass assembly history. 
From the Big Multidark simulation, we extract two large samples of random halos with masses ranging from Virgo to Coma cluster sizes. Additionally, to find the main environmental culprit for the formation history of the Virgo cluster, we compare the Virgo-size halos to 200 Virgo-like halos extracted from simulations that resemble the local Universe.
The number of neighbors at different cluster-centric distances permits discriminating between clusters with different mass accretion histories. Similarly to Virgo-like halos, clusters with numerous neighbors within a distance of about 2 times their virial radius experience a transition at $z\approx1$ between an active period of mass accretion, relative to the mean, and a quiet history. On the contrary, clusters with few neighbors share an opposite trend: from passive to active assembly histories. Additionally, clusters with massive companions within about 4 times their virial radius tend to have recent active merging histories. Therefore, the radial distribution of cluster neighbors provides invaluable insights into the past history of these objects. 

\end{abstract}


\section{Introduction}

Galaxy clusters are the largest virialized objects in the Universe. In the nineties they helped  establishing the concept of a Universe with a matter density below the critical one \citep[see][]{2005RvMP...77..207V} and they played a key role in the development of the current $\Lambda$CDM paradigm \citep[see][for a review]{2011ARA&A..49..409A}. The hierarchical model of structure formation, a key prediction of the $\Lambda$CDM model \citep[e.g.][]{1999MNRAS.308..593C,2002ApJ...573....7E}, can be tested precisely with measuring cluster abundances at different epochs. Observable quantities that are sensitive to the dark matter halo assembly history  can thus contribute to  our  understanding of hierarchical structure formation. Any discrepancies between observations and theory may ultimately point towards necessary modifications of the model including the nature  of the dark matter particle or/and the properties  of the initial density fluctuations \citep[e.g.][]{2009ApJS..180..330K}. Galaxy clusters are thus standard tools for testing cosmological models.

However their utility as probes depends grandly on the control of the various systematic uncertainties and on our understanding of the correlations between observable quantities and their mass.\\

In that respect, cluster structural features, like their fraction of substructures or their mass profile, largely correlated to their formation and evolution have been widely studied \citep[e.g.][]{2008ApJ...682L..73S,2012ApJ...757..102W,2013MNRAS.432.1103L,2013ApJ...763...70W}. The correlation between environment and assembly history has been much less investigated \citep[see e.g.][]{2005MNRAS.362.1099F}. Only a few studies took an interest in studying  potential relations between the large scale environment (i.e. cosmic web)  and the assembly  history of galaxy clusters \citep[e.g][for both observational and theoretical studies]{2017A&A...601A.145F,2018MNRAS.476.4877M}.  A few others focused on the small scale environment but only briefly \citep{2012ApJ...757..102W}. In addition, \citet{2012MNRAS.419.2133H} warn us that the term `environment' is used for a variety of measures that are mostly related to the halo mass. These underlying relations affect the signal that could exist between the current small scale environment of clusters and their assembly history.\\ 

With the advent of larger and larger volume  dark matter only cosmological simulations \citep[e.g.][]{2012MNRAS.426.2046A,2015MNRAS.448.2987F,2016MNRAS.457.4340K} with high enough mass resolution \citep[see e.g][Fig.1 for a review]{2016MNRAS.458..613P}, it becomes  now possible to study the potential small scale - assembly history correlations for a large statistical sample of dark matter halos within a restricted mass range, removing thus the mass dependence. This paper proposes such a study. \\

Beyond looking for a correlation between the current environment of clusters and their accretion histories in general, this paper pursues also a second goal with identifying key properties of our local environment that are responsible for the history of Virgo, our closest cluster-neighbor. Indeed in previous work, using  simulations designed to resemble our local Universe\footnote{The initial conditions of such simulations stem from the $\Lambda$CDM paradigm like any cosmological initial conditions based on this model. They also match a catalog of local observational constraints to result in simulation with the local large scale structure, including Virgo-like halos at $z=0$ thanks to recent improvements.},  we showed that the simulated Virgo-like clusters have had a quiet assembly history within the past seven gigayears while they were more active earlier on \citep{2016MNRAS.460.2015S}. Namely, while the Virgo-like  clusters used to accrete lots of objects in their early stages of evolution, nowadays they still do but to a much smaller extent. In a more recent study \citep{2019MNRAS.486.3951S} that enlarged our previous sample from 15 to 200 Virgo counterparts and increased their resolution by a factor of 3, we found that this kind of assembly history is rare and that this is most probably due to the local environment. This study compared the properties of more than 400 cluster-size random halos to the 200 Virgo-like halos. At $z=0$ only 18\% of the random halos have, besides a similar merging history from $z=4$ to the end, mean radius, velocity dispersion, number of substructures, spin, velocity, concentration and center of mass offset with respect to the spherical center within 3$\sigma$ of Virgo-like halo properties. This correspondence reduces to 0.5\% at 2$\sigma$ and zero at 1$\sigma$. These small rates are due to large-scale environmentally induced properties like the velocity. In addition, $z\approx1$ appears like the redshift of change between the mean assembly history of the Virgo-like halos and that of random halos: from being more active in accreting mass on average than random halos at $z~>~$1, the Virgo-like halos become quieter for $z~<~$1. It is thus of great interest to understand which characterization of the cluster environment can be associated to such a specific assembly history. \\

To investigate this puzzle as well as more broadly potential small scale - assembly history correlations, one needs a large sample of random cluster-size halos. The Big MultiDark simulation (BigMDPL), one of the largest computational volumes of  the MultiDark simulation series using Planck cosmology \citep{2016MNRAS.457.4340K}, provides us with such a sample. We extract from this large cosmological simulation two different cluster catalogs. One with $\sim$~3000 cluster-size halos with masses within [8--10]$\times$~10$^{14}~\hmsun$ and another set of more than 20,000 halos within the mass range [3.7--5.0]$\times$~10$^{14}~\hmsun$. This second set matches our 200 Virgo-like sample with a mean mass of 4.3$\times$10$^{14}\hmsun$ and a standard deviation of 0.66$\times$10$^{14}\hmsun$. We then compare their evolution.  As a consistency check, it is worth mentioning that  \citet{2015AJ....149..171T} published a compilation of the virial masses of nearby clusters.  Assuming Planck cosmology,  the observational mass estimate of the Virgo cluster translates into M$_{vir}~\sim$~4.7$\times$~10$^{14}~\hmsun$, in good agreement with the masses obtained for the Virgo-like halos. \\

This paper starts in Section 2 with a brief description of our 200 Virgo-like halos used as a gauge to determine our environmental property responsible for such an assembly history. Then, it introduces the BigMDPL simulation and describes at length the samples and subsamples of selected cluster-size random halos used to determine the small scale - assembly history correlations. In Section 3, correlations between the assembly history of the cluster halos and their current number of substructures as well as the present number, mass and cluster-centric distance of their neighbors are sought for. In Section 4, we explore the impact of the presence/absence of massive neighbors on assembly histories to match that of Virgo-like halos. Finally, Section 5 summarizes the main finding: a new correlation between the current neighbors of clusters and their accretion history is confirmed independently of their mass.


\section{Simulations and cluster halo (sub)samples}

For our two-goal study two types of dark matter only simulations are required to build the cluster halo (sub)samples: \\
1) a set of constrained cosmological simulations resembling the local Universe that contain a realistic  Virgo-like halo\\
2) a large-volume cosmological simulation which contains a statistically significant number of massive and Virgo-size cluster halos. \\
All these dark matter only simulations are based on the Planck cosmology  \citep[H$_0$=67.77~km~s$^{-1}$~Mpc$^{-1}$, $\Omega_\Lambda$=0.693, $\Omega_m$=0.307, $\sigma_8$=0.823][]{2014A&A...571A..16P}.

\subsection{Constrained simulations and Virgo-like halos}

We use 200 constrained simulations designed to match the large scale structure around the Local Group within a $\sim150$\hMpc\ sphere radius. Local observational data used to constrain the initial conditions of the simulations are distances of galaxies and groups converted into peculiar velocities \citep{2018MNRAS.476.4362S,2016MNRAS.455.2078S} that are bias minimized \citep{2015MNRAS.450.2644S}.  The details of the algorithms and steps to get the simulations are given in \citet{2019MNRAS.486.3951S}.  \\

Built originally for the latter study, each one of the constrained simulations contains a Virgo-like halo at around the position of the observed Virgo cluster by analogy. To minimize the  computing time the zoom-in technique \citep{2001ApJS..137....1B}, implemented in the  \music\ code  \citep{2011MNRAS.415.2101H}, was used to reach an effective resolution of 2048$^3$ particles in the full box (i.e. a dark matter particle mass of $1.2\times$10$^{9}\hmsun$) within a 10~\hMpc\ radius sphere centered on the Virgo-like halos. 

Virgo-like halo properties and assembly histories were obtained with the  {\sc AHF} code combined to the MergerTree algorithm \citep{2009ApJS..182..608K}. Since solely the mass evolution of the Virgo halos is used for comparison with the large sample of dark matters halos in this paper and since this evolution depends only on the cosmological model, the chosen halo finder method used to derive this mass evolution is not critical. \\

These 200 simulated counterparts of the Virgo cluster constitute our first halo sample:
\begin{itemize} 
\item 200 Virgo-like: these halos match the observations extremely well and share similar properties, including the assembly history.  Namely,  the cosmic variance is effectively reduced down to the cluster scale \citep{2016MNRAS.460.2015S,2018MNRAS.478.5199S,2018A&A...614A.102O,2019MNRAS.486.3951S}. The latter investigation of these simulations led to the current search for the environmental properties responsible for the specificities of the Virgo cluster assembly history.\\
\end{itemize}

\subsection{Cluster halos in the MultiDark simulation}

\begin{figure}
\vspace{-1.6cm}
\hspace{-0.5cm}\includegraphics[width=0.55 \textwidth]{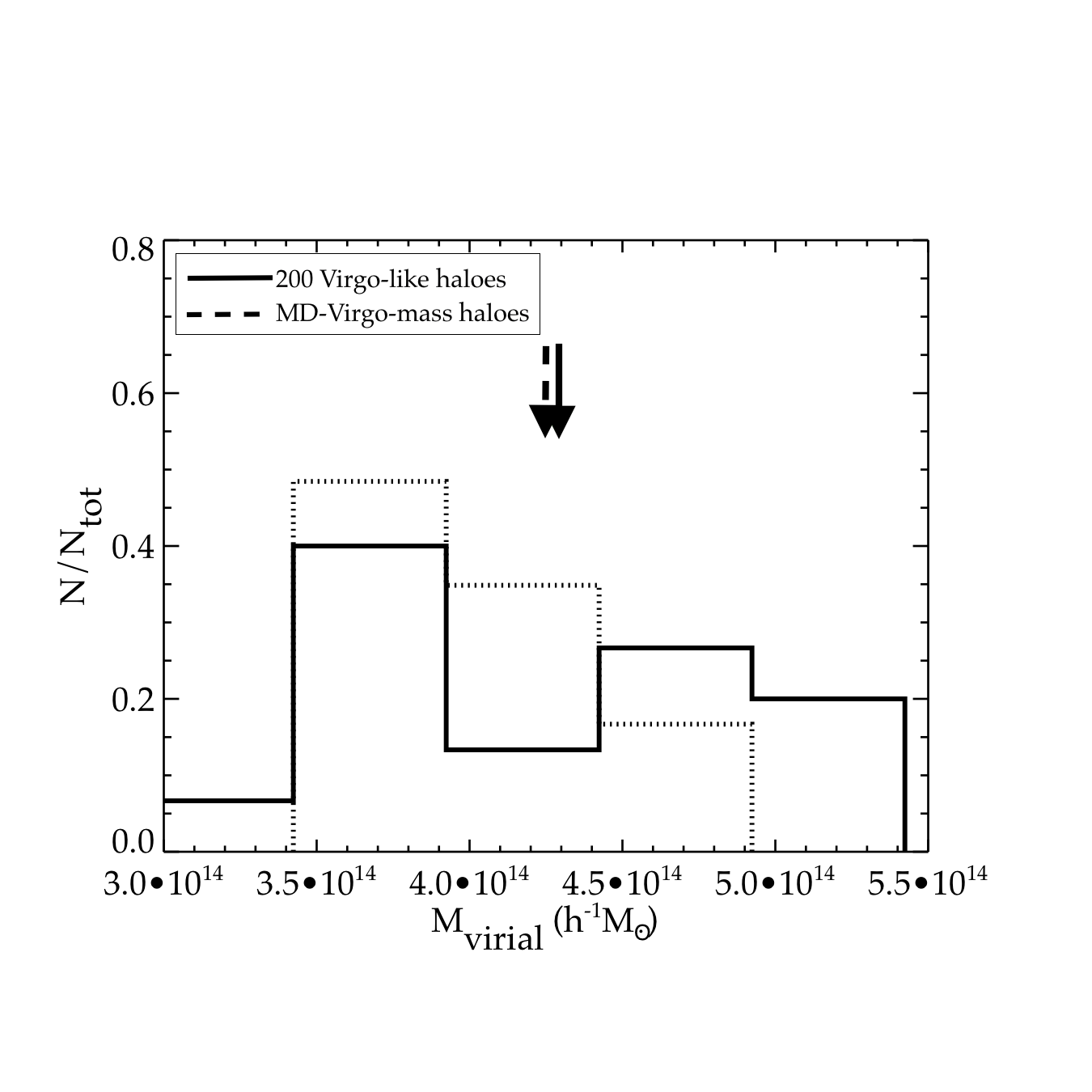}
\vspace{-1.6cm}
\caption{Mass distributions of the 20,226 dark matter halos, selected in the MultiDark simulation to constitute the MD-Virgo-mass sample (dotted line) and of the 200 Virgo-like halos of the constrained simulations (solid line). The arrows stand for the respective mean mass of the Virgo-like and the MD-Virgo-mass halos.}
\label{fig:mass}
\end{figure}

Cluster halos are taken from the BigMDPL simulation which is part of the MultiDark simulation series \citep{2016MNRAS.457.4340K}. It is the second largest boxsize of this series to have been run\footnote{https://www.cosmosim.org/cms/simulations/bigmdpl/}. With 2.5 Gpc~h$^{-1}$ as a side and 3840$^3$ particles, it has a mass resolution  of 2.4$\times$10$^{10}\hmsun$. This simulation has been run with the same cosmological parameters as the constrained simulations. \\

Halo and sub-halo catalogs at different redshifts were extracted from the simulation using the  \rockstar\ algorithm \citep{2013ApJ...762..109B}. Although, all subsequent conclusions are identical when using all halos and sub-halos with more than 50 or 100 particles, to be more conservative, we retain only (sub-)halos with more than a hundred particles. The merger trees were obtained from the \rockstar\  catalogs using the {\sc  consistent trees} software \citep{2013ApJ...763...18B}. From this simulation at $z=0$, we extracted two samples of distinct cluster-size halos that are not substructures of more massive parent halos:
\begin{itemize} 
\item MD-massive: a sample of massive clusters with masses between  [8--10]$\times$10$^{14}~\hmsun$, which contains 2,682 objects at redshift zero;\\
\item MD-Virgo-mass: a sample of Virgo-size halos selected to be within the same mass range as the sample of constrained Virgo-like halos that is described in Section 2.1. More precisely, a halo is retained for further study if its mass is within 1$\sigma$ of the mean mass of the Virgo-like halos where $\sigma^2$ is the mass variance of the Virgo-like halo sample. This second sample contains 20,226 objects. Fig. \ref{fig:mass} shows their mass distribution as well as that of the Virgo-like halos.\\
\end{itemize}

\subsection{Subsamples and deviation from mean assembly history}

Among the three considered sub-samples only the 200 Virgo-like halos share similar environment and assembly history by construction. Instead, the clusters in the MD-massive (respectively MD-Virgo-mass) samples share only the same mass. \\

\begin{figure}
\vspace{-1.cm}
\hspace{-1.cm}\includegraphics[width=0.6 \textwidth]{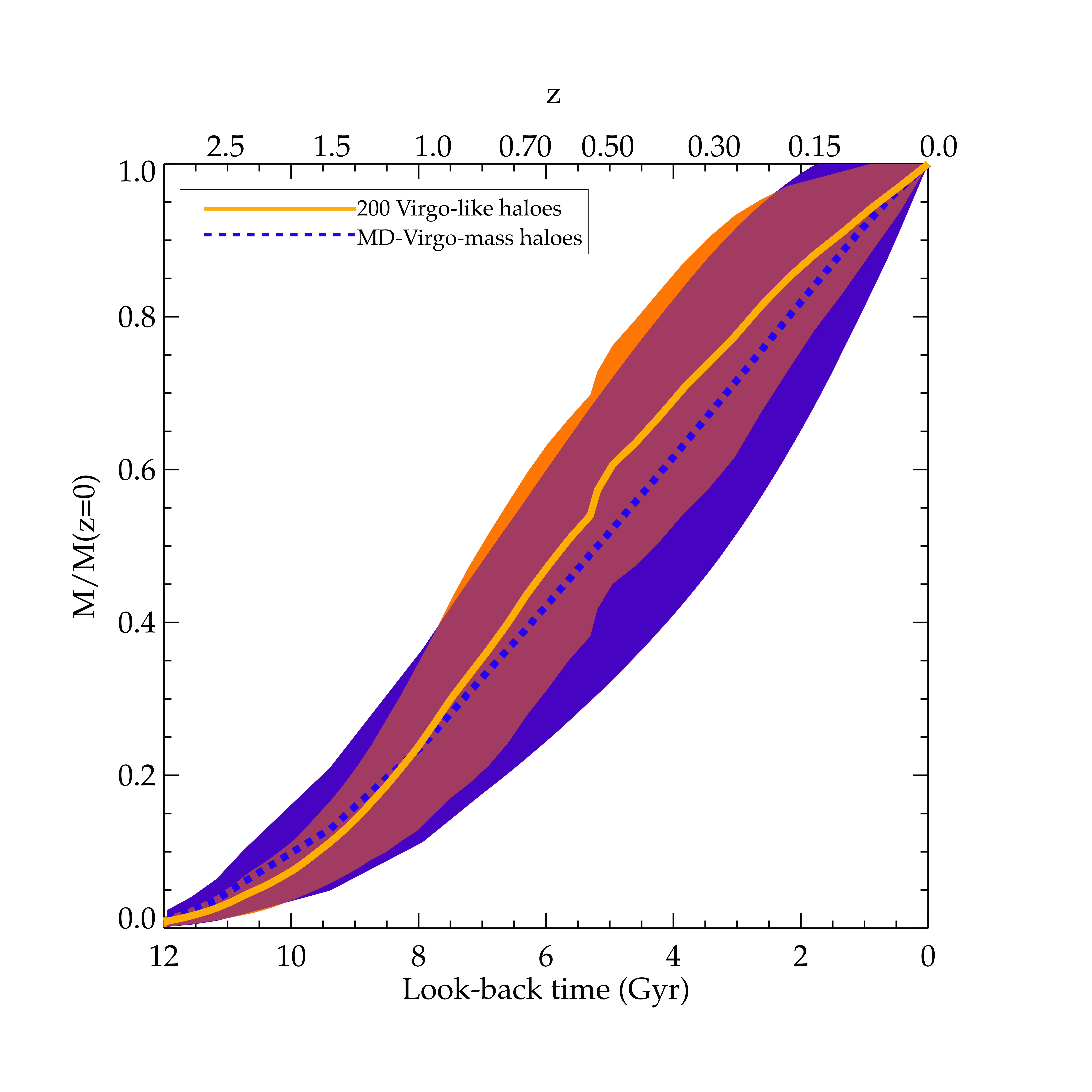}
\vspace{-1.3cm}
\caption{Mean mass assembly history of the MD-Virgo-mass (blue dotted line) and the 200 Virgo-like (orange solid line) samples. Shaded areas give the standard deviations.}
\label{fig:merger}
\end{figure}

As a case in point, Fig. \ref{fig:merger} shows the mean mass assembly history as a function of look-back time (lower axis) or redshift (upper axis)  of 200 Virgo-like (solid orange line) and MD-Virgo-mass (blue dotted line) samples. The colored areas depict their respective standard deviation. As it is clearly shown, the 200 Virgo-halos do not follow the mean of the MD-Virgo-mass halos. In fact, during the last few gigayears, the Virgo-halos had a quieter assembly history than the mean of the random cluster halos.  Our previous papers already demonstrated this result, although based on much smaller random halo samples  \citep{2016MNRAS.460.2015S,2019MNRAS.486.3951S}. Additionally, it is remarkable that despite the much smaller number of halos in the Virgo sample, the scatter of the assembly history of 200 Virgo-like halos is smaller than that of the MD-Virgo-mass sample. This is due to the constrained nature of the simulations.\\

Fig. \ref{fig:merger}  represents also the starting point  of our search for a correlation between environment and mass accretion history. It is indeed striking that constraints set by the observed velocity field of galaxies in a large volume around the local group lead to the prediction of a specific mass assembly history for the Virgo cluster. It is thus interesting to find out whether a subsample of the MD-Virgo-mass halos based on similar environmental properties has an assembly history matching that of the 200 constrained Virgo-like halos. \\

The present numbers of substructures and neighbors are two observable quantities in  clusters. Subsequently, we divide the halo samples into subsamples according to their number of either substructures or neighbors at $z=0$ following Table \ref{tbl:sub} (each subsample range is specified in the tables of Appendix A). The number of substructures/neighbors \textit{per se} is not meaningful: it depends on the resolution of the simulation (i.e. the smallest substructure/neighbor that can be identified) and on the halo finder used. In addition, it is difficult to compare the observed \citep{2014A&A...570A..69B} and the predicted numbers because of projection effects and mass estimate uncertainties in observations especially for substructures. Consequently to be able to apply this study both to simulations and observations, we split the MD-massive and MD-Virgo-mass samples  into different subsamples using the mean number of substructures/neighbors ($\overline{\mathrm n}$) above a given mass and its standard deviation ($\sigma_{\mathrm n}$)  as references. \\

\begin{figure*}

\centering
\includegraphics[width=0.8 \textwidth]{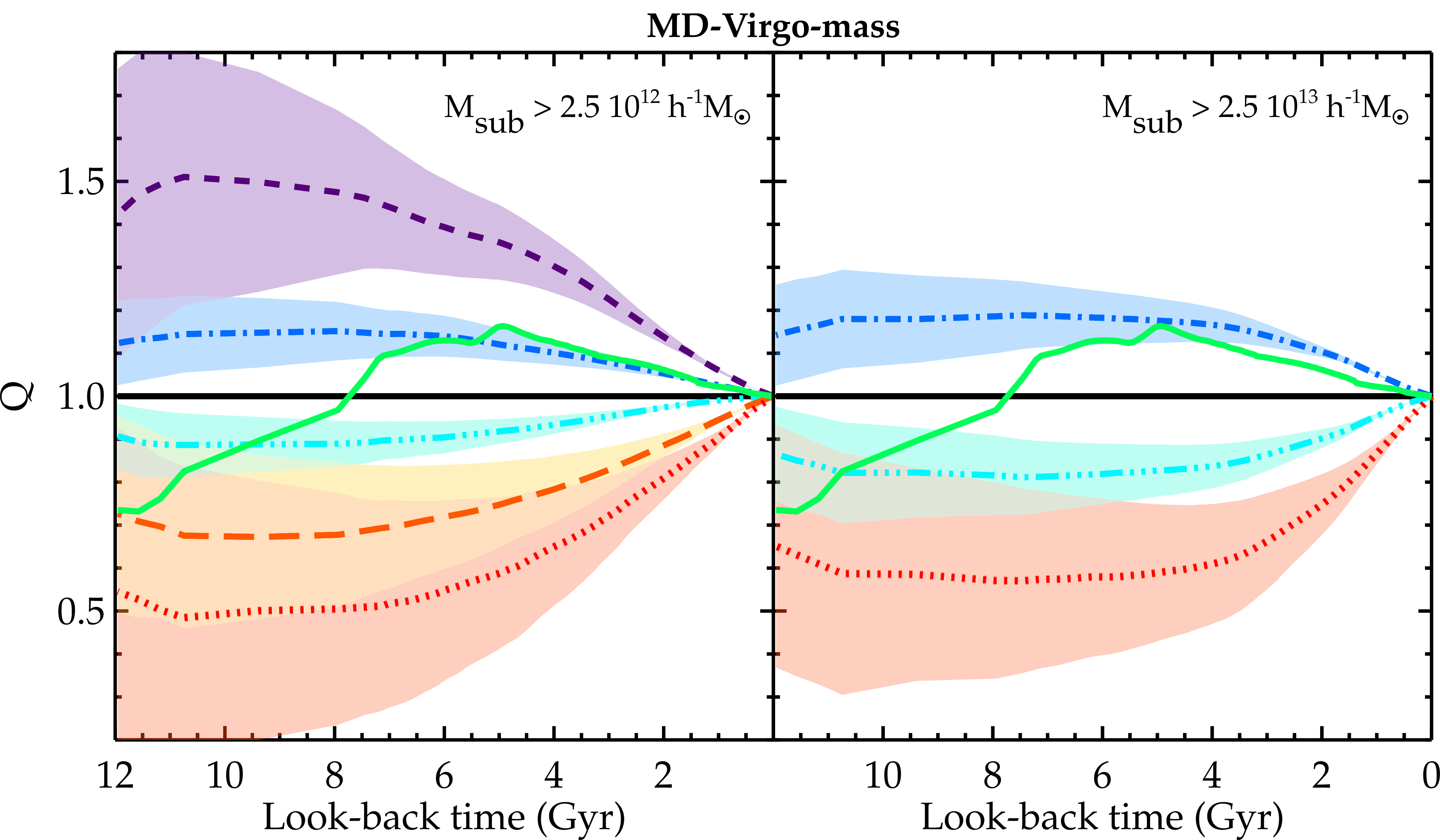}

\vspace{0.5cm}
\includegraphics[width=0.8 \textwidth]{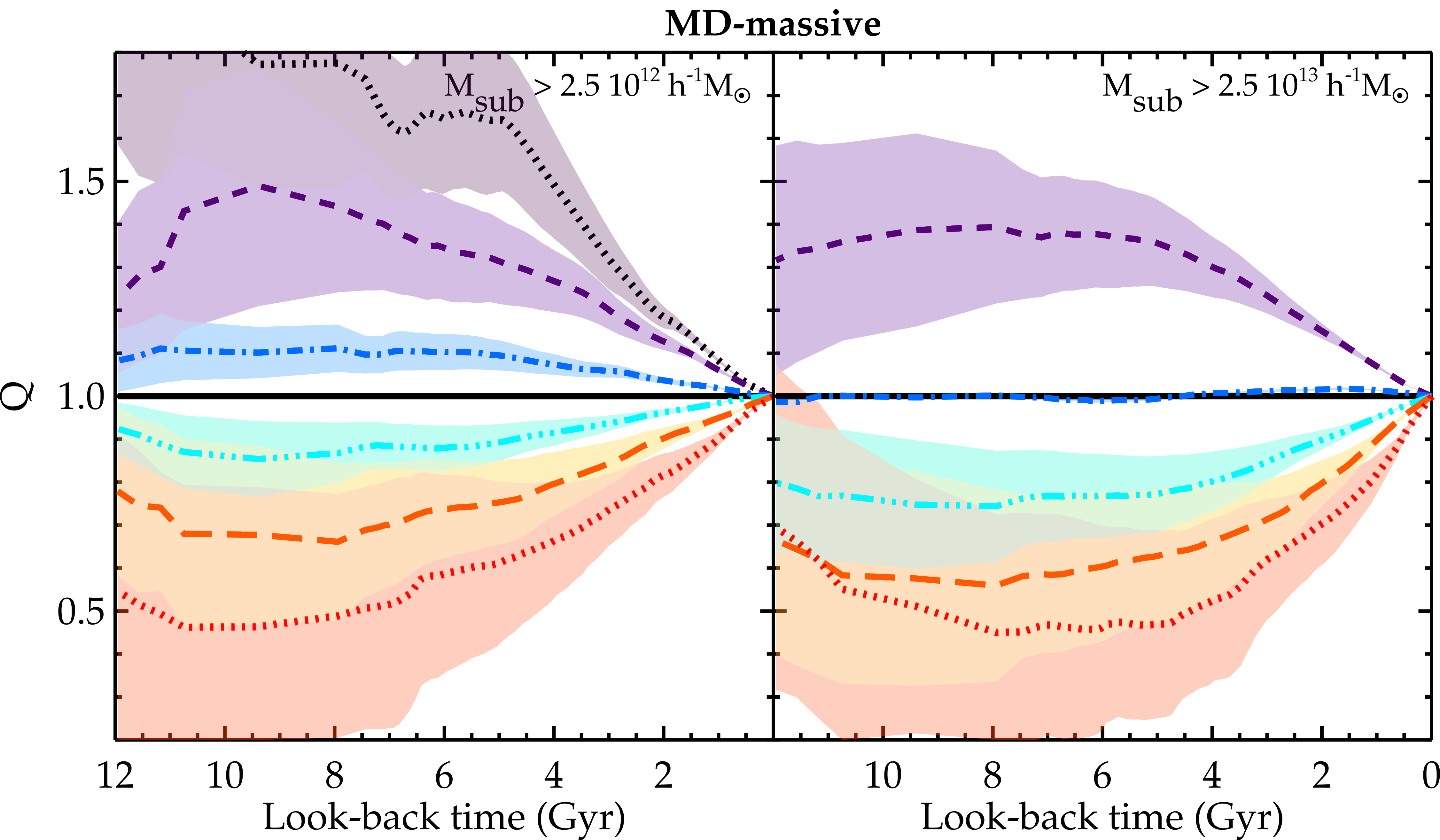}
\caption{$Q$ ratios for the MD-Virgo-mass (top panels) and MD-massive (bottom panels) samples. Left (right) panels are limited to substructures with masses M$_{sub}$ greater than 2.5$\times$10$^{12}\hmsun$ (2.5$\times$10$^{13}\hmsun$). Different colored lines (Table \ref{tbl:sub}) show the trend of  subsamples built on the basis of the number of substructures which is increasing from the black dotted line to the red dotted line (see Table A1). Colored areas give the standard deviations. The solid green lines in the top panels stand for the ratio $Q$ obtained for the 200 Virgo-like halos with respect to the MD-Virgo-mass sample.}
\label{fig:sub}
\end{figure*}

\begin{table}
\begin{center} 
\begin{tabular}{ccc}
\hline
\hline
Sample & line type & color of lines and areas  \\
\hline
\hline
n$~<~\overline{\mathrm n}$~-~2$\sigma_{\mathrm n}$ &dotted    & black line (dark violet) \\
$\overline{\mathrm n}$~-~2$\sigma_{\mathrm n}<~$~n$~<~\overline{\mathrm n}$~-~$\sigma_{\mathrm n}$& short dashed  & violet line (light violet) \\
$\overline{\mathrm n}$~-~$\sigma_{\mathrm n}<~$~n$~<~\overline{\mathrm n}$&dash dotted&blue (light blue)\\
$\overline{\mathrm n}~<~$~n$~<~\overline{\mathrm n}$~+~$\sigma_{\mathrm n}$&dash three dotted & blue-green \\
&  &  (light blue-green) \\
$\overline{\mathrm n}$~+~$\sigma_{\mathrm n}<~$~n$~<~\overline{\mathrm n}$~+~2$\sigma_{\mathrm n}$&  long dashed  &orange (light orange)\\
n$~>~\overline{\mathrm n}$~+~2$\sigma_{\mathrm n}$&dotted &red (light red)  \\
\hline
\hline
\end{tabular}
\end{center}
\vspace{-0.25cm}
\caption{Definition of the different halo subsamples based on their number $n$ of substructures and neighbors. The line types and the colored regions refer to Figs. \ref{fig:sub}, \ref{fig:mmh1} and \ref{fig:mmh2}.}
\label{tbl:sub}
\end{table}

Subsequently, cluster halos can then be categorized from those with no or very few substructures/neighbors up to halos with a large number of substructures/neighbors going through intermediate numbers. The main goal of this paper is to find a new probe that can be used in observational studies. The observational counterparts of simulated dark matter substructures are the galaxies and sub-groups of a cluster that are grandly affected by project effects. Therefore, using substructures for our analyses, while interesting, will not be particularly usable in observations. Observationally, cluster neighbors are more easily identified and characterized than substructures. Thus, we go further by introducing neighbor cluster centric distances and masses in the subsample selection criteria. This can be done similarly in all the simulated and/or observed samples. Furthermore, this process should permit classifying observed clusters into the above categories even in the case of random and/or systematic biases applying to the whole observational sample provided that the latter is statistically significant  (i.e. with significant mean and standard deviation). Appendix A gives the number of halos per subsample as well as the different ranges for the number of substructures/neighbors.\\

In order to find possible correlations between the assembly history of clusters and their \emph{current} cluster properties,  like the number of substructures  or the number, mass and cluster-centric distances of neighbors, we then compare the mean assembly histories of the halos in the different subsamples with the mean merger history of the total samples and of the constrained Virgo halos for the MD-Virgo-mass sample. To that end, we define the ratio $Q(t)$ as follows:
\begin{equation}
Q(t)= \frac{ \frac{1}{N_{c}} \sum\limits_{j=1}^{N_{c}} M_{j, \, \mathrm virial}(t) / M_{j, \, \mathrm virial}(0)}{ \frac{1}{N} \sum\limits_{i=1}^N M_{i, \, \mathrm virial}(t) / M_{i, \, \mathrm virial}(0)}
\label{eq:1}
\end{equation} 
where $N$ is the total number of cluster halos and  $N_{c}$ is the number of cluster halos that match a given criterion $c$ in either MD-Virgo-mass or MD-massive. $M_{i, \,\mathrm  virial}(t)$ is the virial mass of the halo $i$ at look-back time  $t$ and today $t=0$. 

In other words, the quantity $Q$ is the ratio of mean assembly histories or the deviation from the mean assembly history at a given time. Any deviation from the unity means that the halos selected under the criterion $c$ have on average a history that deviates from the mean. If  $Q(t) > 1$ at a given time, the selected halos are quieter  than on average. They  already grew in the past to reach their mass value at present. Reversely, if  $Q(t) < 1$  the selected halos are at that time more active than on average since they need to grow faster to reach their mass today.

\section{Assembly history}
 
Rather than focusing on the formation time of galaxy clusters like in  previous studies  \citep[e.g.][]{2012ApJ...757..102W}, analyses in this paper are directed towards the type of accretion history: passive, active, quiet or a combination of these across cosmic time. These adjectives are used to describe the mass evolution or growth of the halos. A fast growth is associated with an active assembly history while a slow increase in mass is due to a quiet or even passive history in case of quasi or even absence  of matter accretion.

\subsection{Number of substructures}
 
The number of substructures of a halo is defined as the number of subhalos within its virial radius. The virial radius of a halo is defined as the radius of a sphere whose density is $\Delta_{vir}(z)\times\Omega_m\rho_c$ at that redshift. $\Delta_{vir}(z)$ is given by the spherical collapse model in a given cosmology. In \rockstar, this value is taken from the analytical fitting formula given in \citet{1998ApJ...495...80B}.\\

Fig. \ref{fig:sub} presents $Q(t)$: the different mean assembly histories of halo subsamples relative to the total mean history of the MD-Virgo-mass and MD-massive  samples respectively. As expected, clusters with the largest number of substructures (more massive than 2.5$\times$10$^{12}\hmsun$, left panels or more massive than 2.5$\times$10$^{13}\hmsun$, right panels) have had on average the most active assembly histories ($Q(t)<1$, warmest colors) while those with a few substructures have had on average the most passive histories  ($Q(t)>1$, coldest colors). Within the last few gigayears, the former grows faster than the average while the latter grows slower. Light colored areas stand for standard deviations of the mean curves. The different scenarios are quite distinguishable,  confirming  the link between the current number of substructures and the assembly history of galaxy clusters already reported by  e.g. \citet{2012MNRAS.419.3280S}.  \\

However, none of the subsamples built from the MD-Virgo-mass sample presents a change of trend at $z\approx1$ like that observed for the Virgo-like halos (solid green line). In other words, no average line crosses the ordinate equals to 1 line at $z\approx1$ which would indicate that this subsample contains halos that used to grow faster than on average at $z>1$  and that later grow slower than on average. This means that Virgo-like halos cannot be identified solely by the number of substructures they have at present. This is quite expected: while the Virgo halos have had a quiet assembly history within the past few gigayears, they have a larger, rather than a smaller, total number of substructures than random halos on average \citep{2019MNRAS.486.3951S}. The number of substructures is therefore not sufficient to identify halos with a assembly history similar to that of Virgo-like halos. Therefore, in the next section we explore this issue in more detail and focus on the abundance, masses and cluster-centric distances  of cluster neighbors at present. 

\subsection{Number and clustercentric distance of neighbors}

\begin{figure*}
\vspace{-0.75cm}
\includegraphics[width=0.88 \textwidth]{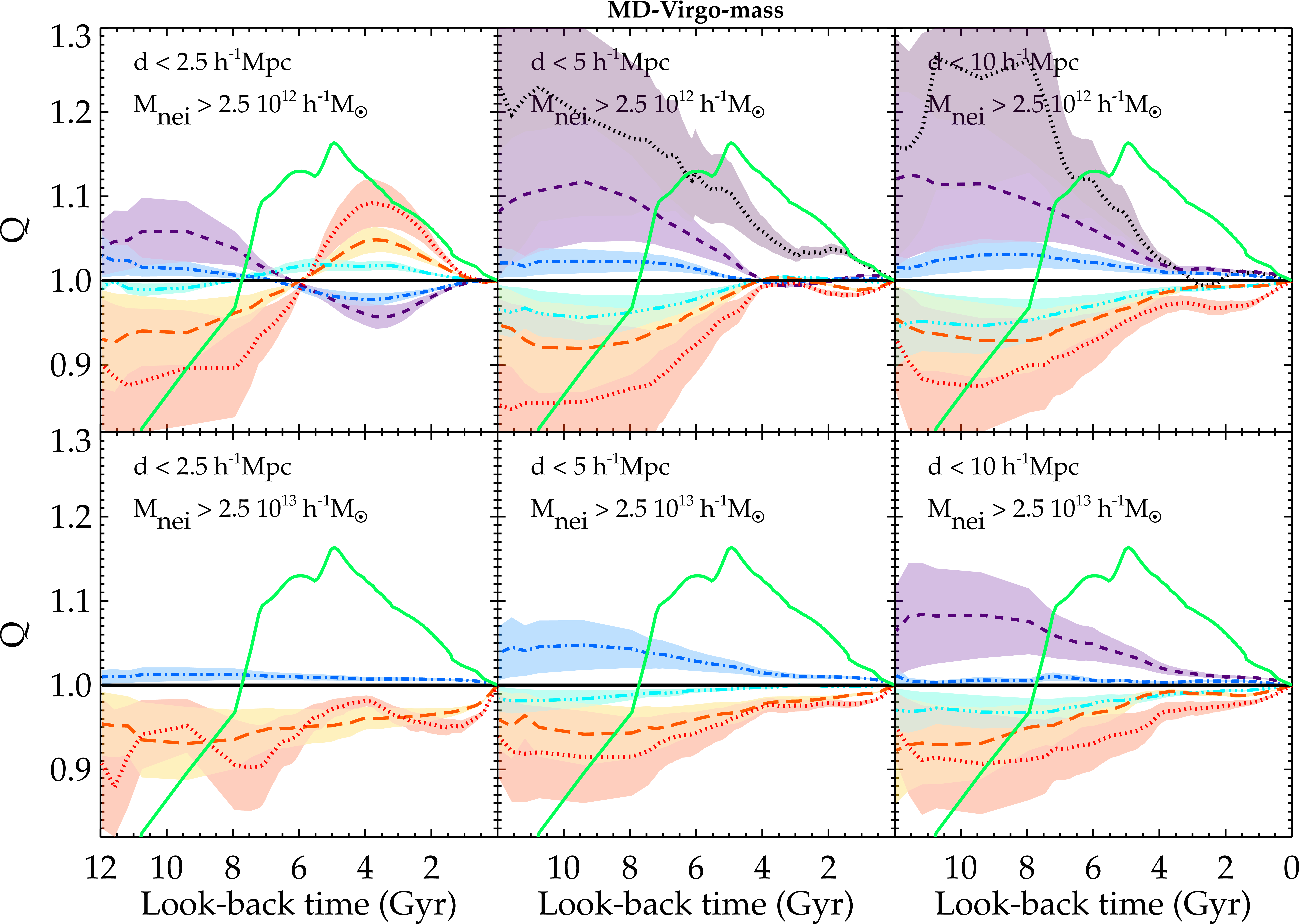}
\vspace{0.5cm}

\includegraphics[width=0.88 \textwidth]{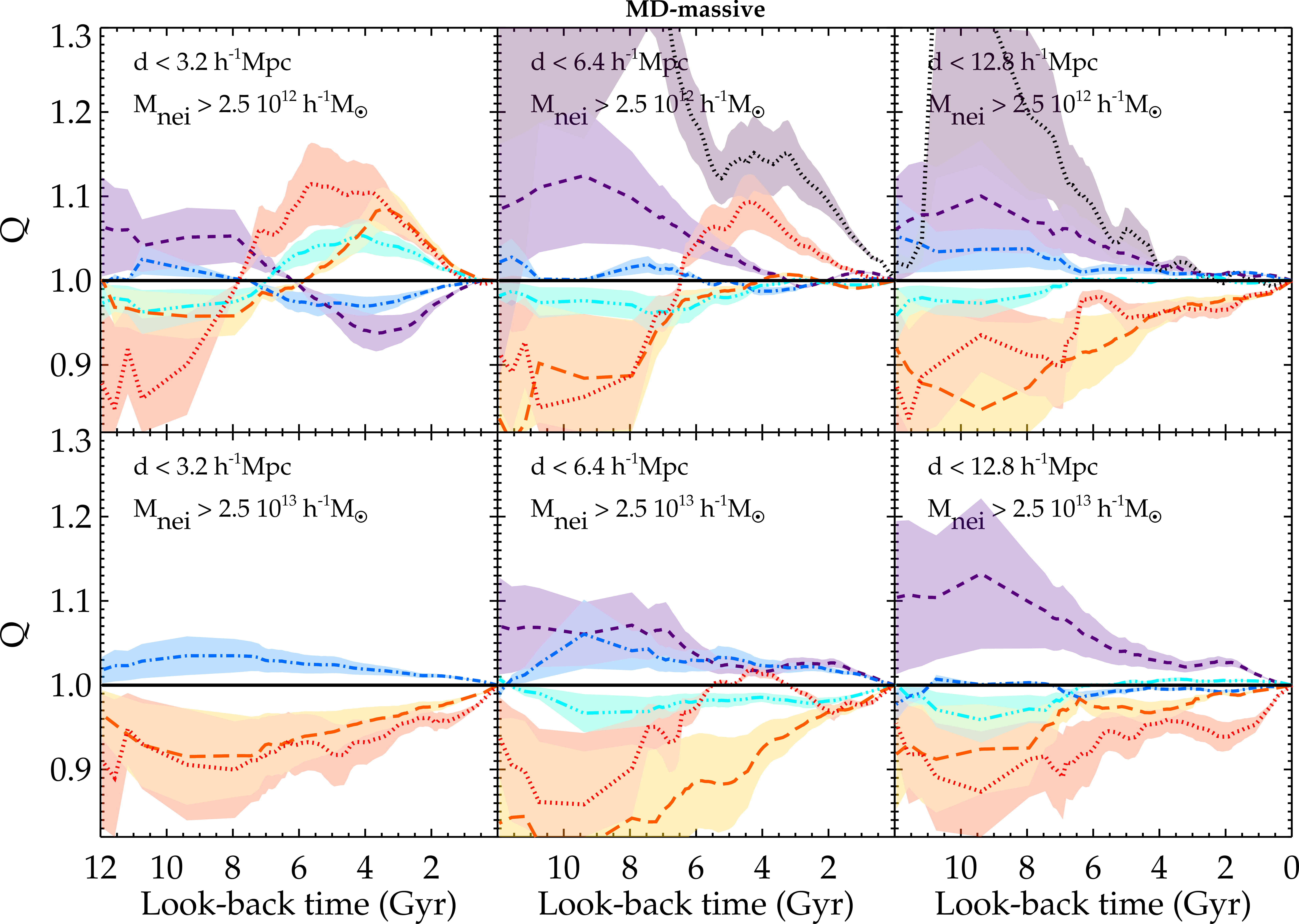}
\vspace{-0.15cm}
\caption{$Q$ ratios for the MD-Virgo-mass (top 6 panels) and MD-massive (bottom 6 panels) samples. First and third (second and fourth) row panels account for neighbors more massive than 2.5$\times$10$^{12}\hmsun$ (2.5$\times$10$^{13}\hmsun$). From left to right, more and more distant (within $\sim2~r_{vir}$, $\sim4~r_{vir}$ and $\sim8~r_{vir}$) neighbors are included in the count. Different colored lines (Table \ref{tbl:sub}) show the trend of  subsamples built on the basis of the present number of neighbors which is increasing from the black dotted line to the red dotted line (see Table A2). Colored areas give the standard deviations. The solid green lines in the two first rows of panels stand for the ratio $Q$ obtained for the 200 Virgo-like halos with respect to the MD-Virgo-mass sample.}
\label{fig:mmh1}
\end{figure*}

In this section the cluster halo samples are split into 6 subsamples according to the number of neighbors within a given distance (from the virial radius of the cluster halos to either $\sim2~r_{vir}$ or  $\sim4~r_{vir}$   or $\sim8~r_{vir}$) and with a given minimum mass (either 2.5$\times$$10^{12} \hmsun$ or 2.5$\times$$10^{13} \hmsun$). In Fig. \ref{fig:mmh1} the ratio $Q(t)$  (Eq. \ref{eq:1}) is shown for the different subsamples of both the MD-Virgo-mass (top) and MD massive (bottom) samples. As in the previous figure, standard deviations are shown as light colored areas. In the top 6 panels, the solid green line stands for the average assembly history of the 200 Virgo-like halos divided by the mean assembly history of all the halos in MD-Virgo-mass. Findings as described below are here again quite similar for both mass ranges.\\

According to Fig. \ref{fig:mmh1}, the various cluster subsamples exhibit quite different behaviors although these behaviors are similar between halos of the two mass ranges. The main results drawn from this figure are summarized below:\\
\begin{itemize}
\item Left panels of the first and third rows: assembly histories are alternatively quiet or active, i.e. $Q(t) - 1$  changes sign over time. The sought for behavior with a redshift of change appears distinctly. The transition redshift ($z \approx 0.7$ or $t \approx 6$~Gyrs) is close to that observed for the Virgo-like halos ($z \approx 1.0 $ or $t \approx 8$~Gyrs).  Namely, after that redshift, halos with currently many neighbors (dotted and long-dashed lines,  warmest colors) had a passive assembly history ($Q>1$) while before $z=0.8$, they tended to have had an active assembly history ($Q<1$). The reverse is true for halos with presently only a few neighbors (coldest colors). Note that for the most massive halos with the largest number of neighbors (left panel of the third row), their redshift of change is even closer to that of the 200 Virgo-like halos. This is probably due to the smaller number of haloes in that subsample with respect to ten times more halos in the other subsamples, with the 200 Virgo-like halo sample having an intermediate number. The trends are indeed smoother the more populated the subsamples are. It flattens the curves and gives an average intermediate redshift of change of $z\approx0.7$. Note that the absence/presence of a massive neighbor is responsible for this shift in the redshift of change. As we will show below, a massive neighbor contributes to maintaining an accretion activity, and thus shifts the change of behavior to later times while its absence permits an earlier change. To summarize, independently of their mass, halos with currently the largest number of neighbors (dotted line) in a very close vicinity (less than $\sim2~r_{vir}$), entered the quieter assembly history ($Q>1$) mode more recently. \\
\item Middle panels of the first and third rows: transition signals are clearly dampened with the exception of the most massive halos with the largest number of neighbors. There are small ripples recently but it is harder to discriminate between the assembly histories of halos when including their more distant neighbors. \\
\item Right panels of the first and third rows: the above mentioned trend is confirmed. The assembly histories of the different subsamples are mostly quiet or active at all times. Namely, when including neighbors within 10-12 \hMpc\  the transition disappears. It makes it more challenging to discriminate the recent assembly histories of halos. This is most probably due to both the increasing probability of encountering more and more massive neighbors with the distance from the clusters and the limit of the `small scale' / cluster interaction.\\ 
\item Left panels of the second and fourth rows: indeed, when considering only massive neighbors, the redshift of transition does not appear anymore. Halos with the largest number of  close-by massive neighbors (dotted line) tend to be more active ($Q<1$) than whose without even at late times. These massive neighbors do not permit discriminating as efficiently as the small neighbors do between halo recent assembly histories. \\
\item Middle and right panels of the second and fourth rows : Again the existence of neighbors more massive than 2.5$\times$10$^{13}\hmsun$, i.e. 1-3\% of the main halo mass, prevents a change of the assembly history irrespective of their distance, i.e. an active/quiet assembly history remain on average active/quiet. The massive neighbors of a halo contribute to its gravitational potential and thus support its active history.
\end{itemize}

To summarize, small neighbor counts in the close vicinity permit discriminating the recent assembly history of halos. Massive neighbors help halos accreting mass. Thus they maintain their accretion activity, most probably resulting in a shift of their redshift of change to later times. However, when considering neighbors at larger distances, the signal is dramatically damped. Therefore, there exists a well established correlation between the current number of close-by small neighbors of a cluster-size halo and its past assembly history that can be further refined when considering the presence or absence of massive neighbors.

\section{Massive neighbors drive the redshift of change}

To go deeper into refining our correlation and identifying the type of neighbors required to get a Virgo-like assembly history, it is worth emphasizing yet again that the change of sign of $(Q(t)-1)$  becomes weaker from left to right in the top row of Fig.  \ref{fig:mmh1}. When including more distant neighbors, the signal is damped because of the increasing probability of encountering heavier neighbors than within shorter distances. Since Virgo is known not to have (numerous) massive neighbors within 5~\hMpc\, and actually no neighbor of the same order of mass, it is interesting to further reduce our full sample of random halos within the same mass range as Virgo halos by excluding those that have massive neighbors in their vicinity. \\

To this end, we redefined the ratio $Q(t)$  in Eq. \ref{eq:1}, to hereafter $Q'(t)$, assuming now that  the denominator is summed over a sample of cluster halos,  excluding those  with neighbors more massive than 10$^{13}\hmsun$ within 2.5 and 5~\hMpc. This reduces the total number of cluster halos in MD-Virgo-mass to 13,285 and 3,108 respectively. \\

\begin{figure*}
\includegraphics[width=0.7 \textwidth]{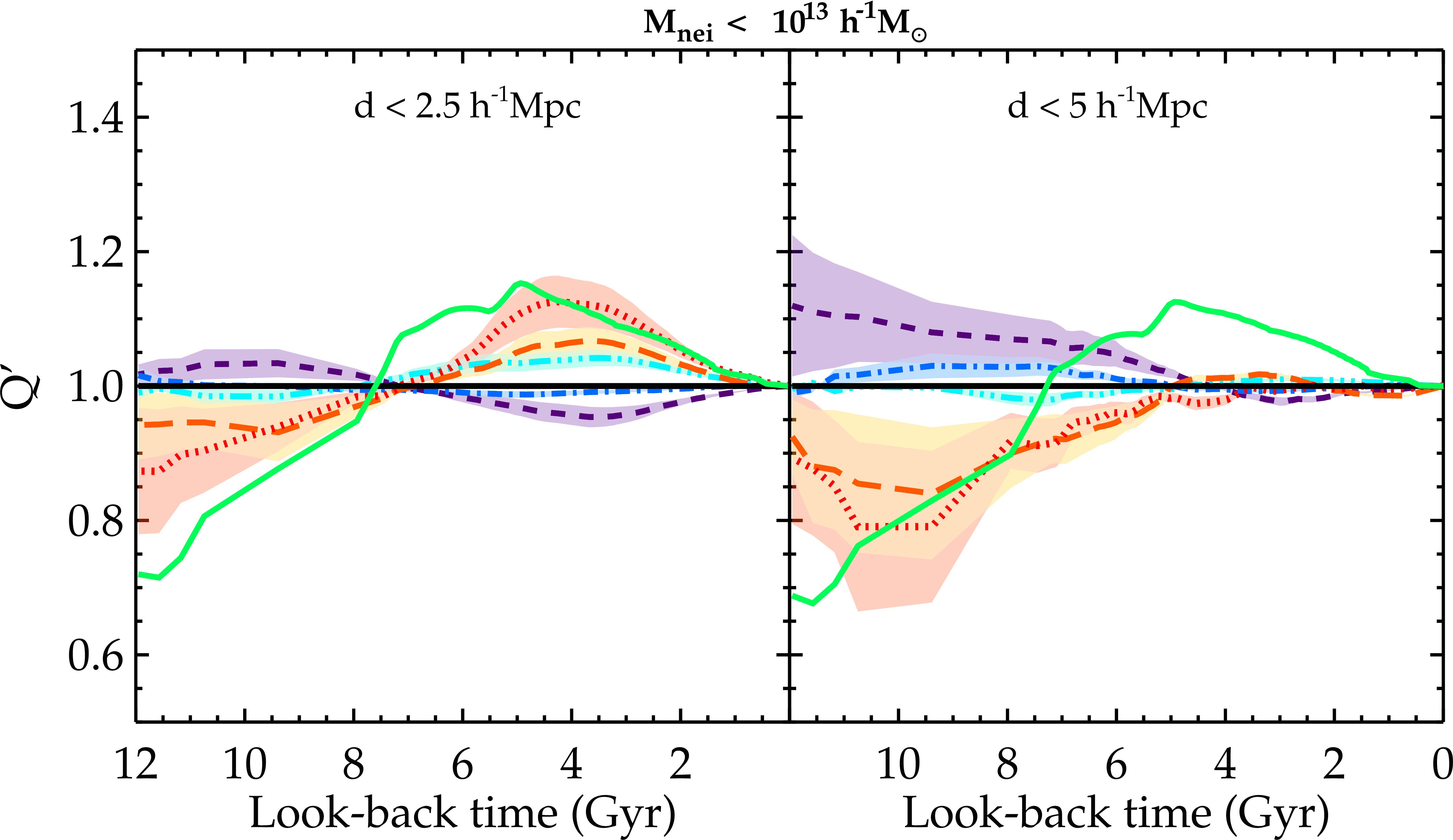}
\caption{Same as the two first panels of Fig. \ref{fig:mmh1} but for a restricted sample of dark matter halos within MD-Virgo-mass: only halos with no neighbors more massive than 10$^{13}\hmsun$ within 2.5  and 5\hMpc\ radii (from left to right) are retained. Different colored lines (Table \ref{tbl:sub}) show the trend of  subsamples built on the basis of the present number of neighbors which is increasing from the black dotted line to the red dotted line (see Table A3). Colored areas give the standard deviations. The solid green lines stand for the ratio $Q$ obtained for the 200 Virgo-like halos with respect to the MD-Virgo-mass restricted sample.}
\label{fig:mmh2}
\end{figure*}

Fig. \ref{fig:mmh2} is thus similar to Fig. \ref{fig:mmh1} but restricted to halos without neighbors more massive than 10$^{13}\hmsun$ within a 2.5 and 5 \hMpc\ radius (from left to right). The left panel shows the best agreement with the mean assembly history of the Virgo counterparts during the last 4 gigayears. The redshift of change is even now shifted to earlier times (from 6 to 7 Gyr, corresponding to a shift from $z\approx0.7$ to $z\approx0.8$).  Note that our sharp limits in distances and masses have a small influence on the results. Typically, allowing a few neighbors slightly above 10$^{13}\hmsun$ within 5~\hMpc\ does not make a significant difference.  We performed several tests to confirm that our conclusions are rather independent on the exact maximum mass value chosen to count neighbors. \\

The fact that the Virgo cluster does not have a nearby massive neighbor, combined with the multitude of leftover small neighbors,  explains its quiet assembly history nowadays as well as its redshift of change. This multitude of close-by small neighbors suggests also that it had an active assembly history in the past. It used to have a strong accretion rate but it slowed down lately. Thus a multitude of small neighbors are still in its  vicinity. They are  approaching  but have not been accreted yet. The environment has not been `wiped' out by accretion unlike for the halos with very few small neighbors: halos that used to be passive and are now active. These halos did not grow in the past (passive history) thus to reach the mean mass of our Virgo sample, they need to accrete mass faster nowadays. This is reinforced by the absence of a massive neighbor that would otherwise maintain its strong accretion rate.  \\

Consequently, we identify the relative number of current neighbors, within $\sim$2.5 to 5\hMpc\ (2 to 4 times the virial radius), with masses at least about 2 orders of magnitude below the mass of the cluster, as the important parameter to determine whether the past assembly history of a Virgo-size halo was more active before and quieter after the redshift of change compared to the average merging history of a large sample of clusters. Observational effects could systematically bias the number of neighbors. This is not critical since this number is not relevant \emph{per se}. This number of neighbors must be compared to the average number of neighbors of the cluster sample. Note that this relation holds for more massive halos. 

A second parameter permits refining more precisely the past history (i.e. the redshift of change): the presence or absence of large neighbors (at least above about 2\% the mass of the cluster) within the same radius. A large neighbor indeed nurtures an active accretion history. \\

This study shows that the Virgo cluster, which has had a quiet assembly history recently while being more active in the past, has an assembly history similar  to that of dark matter halos within the same mass range, without massive companion within 2.5~\hMpc\ but, with a multitude of small neighbors within the same radius. Reversely, halos within the same mass range ,without massive companions but, with very few small neighbors within 2.5~\hMpc\ have had an active assembly history within the last Gigayears and used to be more passive in the past, in the sense that their mass used to evolve slower than an average halo.


Moreover, only 473 out of 20,226 halos present on average a history similar to Virgo, confirming that clusters like our closest neighbor are quite rare. In other words, 13,285 halos of the total sample of 20,226 cluster halos within a 2.5~$h^{-1}$~Gpc cubic volume  have no massive companion within 2.5\hMpc, and only 473 over 20,226 (2.3\%) have a multitude of small neighbors. They have, on average, an assembly history in agreement with that of the Virgo-like cluster lately (see Fig. \ref{fig:mmh2}, left panel). Only a small fraction of  halos share the current environmental properties that imply that they had a similar assembly history as the Virgo-like cluster. A detailed study of these halos with no massive neighbor within 2.5\hMpc\ and lots of small neighbors confirms that they all had a very similar assembly history. Differences are visible only before the redshift of change. In any case, they present the same trend and the variance is smaller than that for all the random halos. We also notice the small variance in terms of the number of small neighbors as a function of the distance from the cluster center. The variance value is at most similar to the variance of the entire sample despite the much higher number of halos in the complete sample than in the selected subsample. \\

The number of current neighbors alongside their mass constitutes thus an alternative to the assembly history type criterion required to select clusters similar to the Virgo cluster in addition to the mass and velocity selection criteria determined  in \citet{2019MNRAS.486.3951S}. \\

\section{Conclusion}

Provided that they are well understood, galaxy clusters are standard tools for testing cosmological models like the hierarchical structure formation of $\Lambda$CDM. Observable quantities that are sensitive to the cluster assembly history constitute thus an inestimable knowledge to compare observed measurements to theoretical expectations.\\

This paper is mainly focused on finding a mass-free correlation between the relative number of current neighbors of cluster-size dark matter halos and their mass assembly history. An underlying additional goal consists in uncovering properties of our environment responsible for the distinctive merging history of our closest cluster neighbor, the Virgo cluster \footnote{Assuming a correlation between the assembly history of clusters and their environment, i.e. halos sharing the same assembly history as the Virgo-like cluster should indeed live in the same environment.}.\\
Indeed, in previous studies based on Virgo-like clusters in the proper large scale environment of constrained simulations, we found that Virgo-like halos have had an active merging history in the past (before $z\approx1$) while they are quieter nowadays (after $z\approx1$) with respect to random halos within the same mass range. \\

To achieve both our goals, we extract from the 2.5~h$^{-1}$Gpc boxsize MultiDark cosmological simulation two cluster size halo samples. The first sample gathers all the halos with masses ranging from $8~\times~$10$^{14}$ to  10$^{15}~\hmsun$ for a total of about 3,000 halos. The second sample is built with halos within the same mass range (within 1 sigma) as our Virgo-like halos for a total of more than 20,000 halos. \\

These large cluster halo samples permit constructing subsamples based on several criteria, in particular with constraints on the current number and masses of neighbors of the cluster halos. Trends arise independently of the halo sample / mass range considered. In fact, halos with currently the largest number of neighbors in their close vicinity have a quieter assembly history recently than on average, while they used to be more active  before $z\approx1$. These halos indeed did not accrete recently the neighbors in their vicinity and thus did not empty their close-by environment. On the contrary, a low number of neighbors in this distance range is linked to the opposite assembly history: recently active and quieter in the past. Finally,  massive companions (mass above about 2\% that of cluster halos)  within 2 to 4~$r_{vir}$ foster recent active assembly histories. \\

The most important parameter to determine the past assembly history of a cluster  is thus the relative number of current neighbors with masses $\sim2$  orders  of magnitude below the mass of the cluster and within 1 to 4~$r_{vir}$. \\

Additionally, determining the presence or absence of massive neighbors (more massive than about $ 1/10^{th}$ the mass of the cluster) within the same range of distances permits refining the selection of halos matching the past history of Virgo-like. If there is no such massive neighbor within 2.5\hMpc, the assembly histories of the Virgo-like clusters  and of the corresponding subsample of cluster halos agree quite well over the most recent cosmic time. A comparison between the small number of halos in the relevant subsamples to those of the large cluster halo sample confirms that the mass assembly history of the Virgo-like cluster (and therefore the Virgo cluster) is rare among all the possible merging histories. About 65\% of the total sample of cluster halos have lots of small neighbors within short distances and less than 4\% have in addition no companion more massive than about 2-3\% of their mass. \\

To conclude, this study confirms that there exists a strong  correlation between the current number of neighbors and the assembly history of clusters independently of the cluster mass. Eventually, it means that the environmental knowledge gives an alternative to the assembly history type criterion required to select clusters similar to the Virgo cluster, a criterion to be added to the mass and velocity selection criteria given in \citet{2019MNRAS.486.3951S}. Because this correlation is based on relative rather than exact numbers of neighbors, it is expected to hold for both higher resolution and hydrodynamical simulations. Indeed the higher resolution simulations will at most permit perhaps pushing the discrimination to higher redshifts by refining subsamples using smaller neighbor masses. Results based on already fully resolved neighbors will however not be affected. As for the hydrodynamical simulations, while the exact number of neighbors might be affected, there is no reason for the relative number to change because the exact numbers will be modified in the same way. This correlation is thus a priori valid for observations.\\

\section*{Acknowledgements}
We would like to thank the referee for their very useful comments that helped clarify the paper. GY acknowledges financial support  from  MINECO/FEDER under research grant AYA2015-63810-P and MICIU/FEDER  PGC2018-094975-C21. The authors would like to thank Adi Nusser for very useful discussions and suggestions. The authors gratefully acknowledge the Gauss Centre for Supercomputing e.V. (www.gauss-centre.eu) for funding this project by providing computing time on the GCS Supercomputer SuperMUC at Leibniz Supercomputing Centre (www.lrz.de). The \rockstar\ catalogs and trees are available at the CosmoSim database (www.cosmosim.org) which is a service by the Leibniz-Institute for Astrophysics Potsdam (AIP).

\section*{Data availability}
Two types of simulations are used in this paper. All the data regarding the random simulation are available at https://www.cosmosim.org/cms/simulations/bigmdpl/ . As for the information regarding the Virgo cluster from constrained simulations, they are available upon request to the authors.


\bibliographystyle{mn2e}

\bibliography{biblicomplete}

\appendix

\section{Exact number of substructures and neighbors of the subsamples}



Tables A1, A2 and A3 summarize the ranges of the number of substructures/neighbors ($n$) that defines each subsample as well as the number of halos ($n_{halo}$) in each one of them. For empty subsamples, the corresponding entry in the table and the line in the associated figure are absent.

\begin{table}
\begin{centering}
\begin{tabular}{|cr||cr|}
\hline
 \multicolumn{2}{|c||}{Left panel} &  \multicolumn{2}{c|}{Right panel}\\
n & n$_{halo}$&  n & n$_{halo}$\\
 & & & \\
\lbrack  0, 1.78\rbrack   & 2382    &  & \\
\lbrack  1.78, 3.88\rbrack   & 7062    &     0           & 11523\\
\lbrack  3.88, 5.98\rbrack   & 6559    & \rbrack  0, 1.23\rbrack   & 6807\\
\lbrack  5.98, 8.08\rbrack   & 3719    &  & \\
\lbrack  8.08,+$\infty$\lbrack   & 504 & \lbrack  1.23,+$\infty$\lbrack   & 1896 \\
                   &20226   &                    &20226 \\
\hline
\end{tabular}

\begin{tabular}{|cr||cr|}
\hline
 \multicolumn{2}{|c||}{Left panel} &  \multicolumn{2}{c|}{Right panel}\\
n & n$_{halo}$&  n & n$_{halo}$\\
\lbrack 0, 1.55\rbrack       & 24       &  & \\
\lbrack  1.55, 4.61\rbrack   & 397  &  0          & 752\\
\lbrack  4.61, 7.66\rbrack   & 918  &\rbrack  0, 1.19\rbrack   & 1026 \\
\lbrack  7.66, 10.7\rbrack   & 879  &\lbrack  1.19, 2.21\rbrack   & 614 \\
\lbrack  10.7, 13.7\rbrack   & 358  &\lbrack  2.21, 3.23\rbrack   & 234 \\
\lbrack  13.7,+$\infty$\lbrack   & 106 &\lbrack  3.23,+$\infty$\lbrack   & 56\\
                   &2682           &        &2682\\
\hline
\end{tabular}
\caption{$n$: range of substructure numbers,  $n_{halo}$: number of halos per subsample for the 4 panels in Fig. \ref{fig:sub}. For empty subsamples, the corresponding entry in the table and the line in the figure are absent.}
\end{centering}
\label{tbl:sub1}
\end{table}

\begin{table}
\begin{centering}

\begin{tabular}{|cr||cr||cr|}
\hline
 \multicolumn{2}{|c||}{Top left panel} &  \multicolumn{2}{c||}{Top middle panel}& \multicolumn{2}{c|}{Top right panel}\\
n & n$_{halo}$&  n & n$_{halo}$ & n & n$_{halo}$\\
 &   &                    0     & 97 &\lbrack 0, 8.07\rbrack       & 145\\
 0           & 4325  &\rbrack  0, 3.43\rbrack   & 3094 &\lbrack  8.07, 16.0\rbrack   & 3297\\
\rbrack  0, 1.60\rbrack   & 6453 &\lbrack  3.43, 6.45\rbrack   & 7881 &\lbrack  16.0, 23.9\rbrack   & 7015\\
\lbrack  1.60, 2.92\rbrack   & 4988 &\lbrack  6.45, 9.47\rbrack   & 6086 &\lbrack  23.9, 31.8\rbrack   & 6445\\
\lbrack  2.92, 4.25\rbrack   & 3847 &\lbrack  9.47, 12.4\rbrack   & 2327 &\lbrack  31.8, 39.8\rbrack   & 2524\\
\lbrack  4.25,+$\infty$\lbrack   & 613 &\lbrack  12.4,+$\infty$\lbrack   & 741 &\lbrack  39.8,+$\infty$\lbrack   & 800\\
                   &20226 &                    &20226 &                    &20226\\
\hline
 \multicolumn{2}{|c||}{Bottom left panel} &  \multicolumn{2}{c||}{Bottom middle panel}& \multicolumn{2}{c|}{Bottom right panel}\\
n & n$_{halo}$&  n & n$_{halo}$ & n & n$_{halo}$\\
&&& &&\\
& &&&\lbrack  0, 1.17\rbrack   & 4108\\
 0           & 16926  &  0            &8565 &\lbrack  1.17, 3.29\rbrack   & 7916\\
 &  &\rbrack  0, 1.85\rbrack   & 7212 &\lbrack  3.29, 5.41\rbrack   & 5316\\
\rbrack  0, 1.03\rbrack   & 2992 &\lbrack  1.85, 2.81\rbrack   & 3149 &\lbrack  5.41, 7.54\rbrack   & 2021\\
\lbrack  1.03,+$\infty$\lbrack   & 308  &\lbrack  2.81,+$\infty$\lbrack   & 1300 &\lbrack  7.54,+$\infty$\lbrack   & 865\\
                   &20226 &                    &20226 &                   &20226\\
\hline
\end{tabular}

\begin{tabular}{|cr||cr||cr|}
\hline
 \multicolumn{2}{|c||}{Top left panel} &  \multicolumn{2}{c||}{Top middle panel}& \multicolumn{2}{c|}{Top right panel}\\
n & n$_{halo}$&  n & n$_{halo}$ & n & n$_{halo}$\\
 &&\lbrack 0, 3.71\rbrack       & 16 &\lbrack 0, 22.4\rbrack       & 27\\
\lbrack  0, 1.24\rbrack   & 525 &\lbrack  3.71, 8.23\rbrack   & 456 &\lbrack  22.4, 34.6\rbrack   & 381\\
\lbrack  1.24, 3.22\rbrack   & 1075 &\lbrack  8.23, 12.7\rbrack   & 899 &\lbrack  34.6, 46.8\rbrack   & 1003\\
\lbrack  3.22, 5.20\rbrack   & 745 &\lbrack  12.7, 17.2\rbrack   & 955 &\lbrack  46.8, 58.9\rbrack   & 821\\
\lbrack  5.20, 7.19\rbrack   & 257 &\lbrack  17.2, 21.7\rbrack   & 252 &\lbrack  58.9, 71.1\rbrack   & 362\\
\lbrack  7.19,+$\infty$\lbrack   & 80 &\lbrack  21.7,+$\infty$\lbrack   & 104 &\lbrack  71.1,+$\infty$\lbrack   & 88\\
                   &2682 &                    &2682 &     &2682\\    
 \hline
\multicolumn{2}{|c||}{Bottom left panel} &  \multicolumn{2}{c||}{Bottom middle panel}& \multicolumn{2}{c|}{Bottom right panel}\\
n & n$_{halo}$&  n & n$_{halo}$ & n & n$_{halo}$\\
&&&&&\\
&& 0           & 500 &\lbrack  0, 3.13\rbrack   & 501\\
 0   & 1903 &\rbrack  0, 1.72\rbrack   & 839 &\lbrack  3.13, 6.26\rbrack   & 1046\\
 & &\lbrack  1.72, 3.08\rbrack   & 1076 &\lbrack  6.26, 9.39\rbrack   & 742\\
\rbrack  0, 1.58\rbrack   & 636 &\lbrack  3.08, 4.44\rbrack   & 172 &\lbrack  9.39, 12.5\rbrack   & 282\\
\lbrack  1.58,+$\infty$\lbrack   & 143 &\lbrack  4.44,+$\infty$\lbrack   & 95 &\lbrack  12.5,+$\infty$\lbrack   & 111\\
                   &2682 &    &2682 & &2682\\
\hline
\end{tabular}
\caption{$n$: range of neighbor numbers,  $n_{halo}$: number of halos per subsample for the 12 panels in Fig. \ref{fig:mmh1}.}
\end{centering}
\label{tbl:mmh1}
\end{table}

\begin{table}
\begin{centering}

\begin{tabular}{|cr||cr|}
\hline
 \multicolumn{2}{|c||}{Left panel} &  \multicolumn{2}{c|}{Right panel}\\
n & n$_{halo}$&  n & n$_{halo}$ \\
 0           & 4325 & \lbrack  0, 1.73\rbrack   & 377\\
\lbrack  0, 1.16\rbrack   & 4703 &\lbrack  1.73, 3.87\rbrack   & 1080\\
\lbrack  1.16, 2.28\rbrack   & 2671  &\lbrack  3.87, 6.01\rbrack   & 1291\\
\lbrack  2.28, 3.40\rbrack   & 1113 &\lbrack  6.01, 8.15\rbrack   & 281\\
\lbrack  3.40,+$\infty$\lbrack   & 473 &\lbrack  8.15,+$\infty$\lbrack   & 79\\
                   &13285 &                    &3108\\
\hline
\end{tabular}
\caption{$n$: range of neighbor numbers,  $n_{halo}$: number of halos per subsample for the 2 panels in Fig. \ref{fig:mmh2}.}
\end{centering}
\label{tbl:mmh2}
\end{table}

\vspace{0.25cm}

 \label{lastpage}
\end{document}